\documentclass[aps,prc,twocolumn,groupedaddress,showpacs,preprintnumbers]{revtex4}

\usepackage[dvips]{graphicx}
\usepackage[dvips]{hyperref}
\hypersetup{pdfborder={0 0 0},colorlinks=true,linkcolor=blue,citecolor=blue,linkbordercolor={0 0 0}}

\newcommand{\rhob}{\bar{\rho}}
\newcommand{\Li}{\text{Li}}

\newcommand{\rhomb}{\rho_{_{\text{MB}}}}

\begin{document}

\title{
Level density of a Fermi gas and integer partitions: a Gumbel-like finite-size correction
}

\author{J\'er\^ome Roccia$^{1}$ and Patricio Leboeuf$\, ^2$}

\affiliation{
$^1$Institut de Physique et Chimie des Mat{\'e}riaux de Strasbourg,
UMR 7504, CNRS-UdS,
23 rue du Loess, BP 43, 67034 Strasbourg Cedex 2, France\\
$^2$Laboratoire de Physique Th\'eorique
et Mod\`eles Statistiques, CNRS,
Universit\'e Paris Sud, UMR 8626, 91405 Orsay Cedex, France}

\begin{abstract} 
We investigate the many-body level density of gas of non-interacting fermions. We determine its 
behavior as a function of the temperature and the number of particles. As the temperature increases,
and beyond the usual Sommerfeld expansion that describes the degenerate gas behavior, corrections due 
to a finite number of particles lead to Gumbel-like contributions. We discuss connections with the 
partition problem in number theory, extreme value statistics as well as differences with respect to 
the Bose gas.
\end{abstract} 


%%% PACS %%%
%%%
%%% 03.65.Sq : semiclassical theories and applications
%%%
%%% 03.75.Ss : degenerate Fermi gases
%%%
%%% 21.10.Ma : Level density 
%%%
%%% 05.30.Fk : Fermion systems and electron gas
%%%
%%% 71.10.-w : theories and models of many electron systems
%%%
%%% 05.30.-d : Quantum statistical mechanics
%%%
%%% 05.40.-a : Fluctuation phenomena, random processes, noise, and Brownian motion 
%%%
%%% 02.10.De : Algebraic structures and number theory 

%\date{\today}

\maketitle

%------------------------------------------------------------------------------------------
%------------------------------------------------------------------------------------------
%------------------------------------------------------------------------------------------
%------------------------------------------------------------------------------------------

\section{Introduction}\label{intro}
The many-body (MB) level density $\rhomb$ is a fundamental quantity which describes basic properties of thermodynamical systems. Its behavior is of principal interest in many different fields such as nuclear decay rate \cite{nuc1,nuc2}, thermonuclear reactions in stellar processes or - more recently - black hole entropy \cite{bh1,bh2,bh3}. Unfortunately {\it ab initio} computations are out of reach  for interacting systems and for large particle numbers. However a statistical treatment is still possible when interactions are taken into account in a mean field approximation. 
For fermions, two extreme regimes can be identified. At temperatures which are low compared to the Fermi energy
$\mu_0$ but high with respect to the mean level spacing at $\mu_0$ (degenerate gas approximation), $\rhomb$ has
a stretched exponential growth \cite{bethe}
\begin{equation}\label{eq_bethe_formul}
\rhomb \propto e^{2\sqrt{\pi^2 \rhob(\mu_0)Q/6}}  \ , 
\end{equation}
where $Q$ is the excitation energy and $\rhob$ is the smooth single-particle (SP) level density. Nowadays Bethe's theory still remains a basic ingredient for advanced MB theories like back-shifted Fermi gas models \cite{dsvu} and the Gilbert-Cameron model \cite{gc}. From Eq.(\ref{eq_bethe_formul}), we see $\rhomb$ is essentially sensitive to the SP level density around the Fermi energy. 
The second regime corresponds to high temperatures $Q\gg \mu_0$, where the classical limit is reached and where $\rhomb$ has a power-law behavior (Maxwell-Boltzmann regime). 
The aim of the article is to analyze, for fermions, the transition between the two regimes (see Fig.\ref{fig}). To tackle this, 
we compute - in the degenerate gas limit - the effect of a finite number of particles when the temperature 
increases. For a power-law SP level density $\rhob(\mu_0)$, we find two contributions to $\rhomb$: the first one 
is given by the usual Sommerfeld expansion while the second one is related to low-lying level excitations. 
This work is in connection with similar results obtained by Comtet {\it et al} for bosons \cite{clm}.
In a mean-field approximation, $\rhomb$ is also directly related to a combinatorial problem: to find the number of configurations of the SP occupied levels for a given excitation energy \cite{tmb,leboeuf}.\\
The paper is organized as follows. In Sec.\ref{gf} we introduce basic definitions of the MB level density. We calculate the expression of the smooth part of the MB level density in the low- and high-temperature regimes. We establish a relation between the partition problem in number theory and the MB level density by using an equidistant spectrum in Sec.\ref{1dho}. We compute corrections due to a finite number of particles from an exact asymptotic formula. Finally we generalize in Sec.\ref{gc} the result to a power-law spectrum using a {\it non-local} Sommerfeld expansion provided by Garoni {\it et al.} \cite{gfg}.
%------------------------------------------------------------------------------------------
%------------------------------------------------------------------------------------------
%------------------------------------------------------------------------------------------
\section{General Framework}\label{gf}
The MB level density is a discrete quantity which, for a gas in a SP potential, is given by:     
\begin{equation} \label{eq_def_rho_MB}
\rhomb(E,N)=\sum_{\lbrace m_j \rbrace} \delta \bigg( E-\sum_{j=0}^{\infty} m_j
\epsilon_j\bigg) 
\delta\bigg(N-\sum_{j=0}^{\infty} m_j\bigg)  \ . 
\end{equation}
It is a microcanonical quantity where the particle number $N$ and the energy of the system $E$ are fixed (all quantities are dimensionless). $\epsilon_j$ corresponds to the energy of the $j$th level of the SP potential for one particle and $m_j$ is the occupation number of the $j$th SP level for a given configuration $\lbrace m_j \rbrace$. In the case of fermions without spin degeneracy, the Pauli principle leads to $m_j$ being either $0$ or $1$. We define $Q$ as the difference $E-E_0$, where $E_0$ is the ground state energy. For each configuration $\lbrace m_j \rbrace$, the conservation of particle number and energy yields ${\sum_{j=0}^{\infty} m_j=N}$ and ${\sum_{j=0}^{\infty} \epsilon_j m_j=E}$, respectively.\\
Now we write Eq.(\ref{eq_def_rho_MB}) in term of an inverse Laplace transform:
\begin{equation}\label{eq_def_transfo_Laplace}
\rhomb(E,N)=\frac{1}{(2 \pi i)^2} \int_{c-i \infty}^{c+i \infty}
\int_{b-i \infty}^{b+i \infty}  e^{\cal{S}(\beta,\mu)} d\beta d\mu \ ,
\end{equation}
where 
\begin{equation}\label{eq_def_entrop}
\mathrm{\cal S}(\beta,\mu)= \beta \bigg[ - \Omega(\beta,\mu) + E -\mu N  \bigg]   
\end{equation}
is the entropy. The constant $b$ (resp. $c$) is chosen such that it is larger (resp. smaller) than the sum index of $\exp[-\beta \Omega(\beta,\mu)]$ with respect to the variable $\beta$ (resp. $\mu$).
Both parameters $\beta$ and $\mu$ correspond respectively to the inverse temperature and to the chemical potential. $\Omega(\beta,\mu)$ is the Grand potential defined as
\begin{equation}\label{def_grd_pot_bos}
\Omega(\beta,\mu) = - \frac{1}{\beta} \int_0^{\infty} 
\rho(\epsilon) \ln \bigg[ 1+ e^{\beta (\mu -\epsilon)}
\bigg ] d\epsilon   \ , 
\end{equation}
where $\rho(\epsilon)=\sum_{j=0}^{\infty} \delta(\epsilon-\epsilon_j)$ is the SP level density.
The saddle point method with respect to $\mu$ and $\beta$ gives the following equations:
\begin{eqnarray} 
\rhomb(E,N)&= &\frac{e^{\cal S (\beta,\mu)}}{2 \pi \sqrt{|\cal D
(\beta,\mu)|}} \ , \label{eq_syst1_rho} \\
N&=&\int_0^{\infty} \frac{\rho (\epsilon)}{
e^{\beta(\epsilon-\mu)} + 1 }\label{eq_syst1_N} \ ,\\ 
E&=&\int_0^{\infty} \frac{\epsilon \rho (\epsilon)}{
e^{\beta(\epsilon-\mu)} + 1 } \label{eq_syst1_E}  \ . 
\end{eqnarray}
$\cal D(\beta,\mu)$ is the determinant of the second derivatives of ${\cal S (\beta,\mu)}$ which is not relevant for the following discussion. The last two equations define $\mu(E,N)$ and $\beta(E,N)$. From Eq.(\ref{eq_syst1_N}), we see that $\mu$ must decrease from positive to negative values when the temperature increases.
Ignoring the discreteness of the SP level density, we now replace $\rho$ by a smooth function:
\begin{equation}\label{eq_rho_power_law}
\rho(\epsilon) \approx \rhob(\epsilon)=\nu \epsilon^{\nu-1}~,~\text{for}~\nu>0 \ .
\end{equation}
For instance, the case $\nu=D$ is the leading order of the $D$-dimensional HO. The case of billiards is given by $\nu=D/2$, where $D$ is the dimension of space. When $\rho$ is given by Eq.(\ref{eq_rho_power_law}),
equations Eqs.(\ref{eq_syst1_N}) and (\ref{eq_syst1_E}) become:
\begin{eqnarray} 
N&=&- \frac{\Gamma (\nu+1)}{\beta^{\nu}} \Li_{\nu} (- e^{\beta
\mu})\label{eq_syst2_N} \ , \\ 
E&=&- \frac{\nu \Gamma (\nu+1)}{\beta^{\nu+1}} \Li_{\nu+1} (-
e^{\beta \mu})  \ , \label{eq_syst2_E} 
\end{eqnarray}
where $\Li_{\nu}(x)= \sum_{k=1}^{\infty} x^k/k^{\nu} $ is the polylogarithm function. In this case the total energy and the Grand potential are connected by $E=-\nu \Omega(\beta,\mu)$. From Eq.(\ref{eq_def_entrop}), the entropy simplifies into
\begin{equation}\label{eq_def_entrop2}
\mathrm{\cal S} = \beta \bigg [ (1+1/\nu) E - \mu N \bigg] \ .
\end{equation}
For the low temperature regime, we expand Eqs.(\ref{eq_syst1_N}) and (\ref{eq_syst1_E}) with $1/(\beta \mu)$ as a small parameter. This so-called Sommerfeld expansion \cite{s,am} is valid for any regular function $f$:
\begin{equation}\label{eq_dev_sommerfeld}
\int_0^{\infty} \frac{f(\epsilon) d\epsilon}{e^{\beta(\epsilon-\mu)}+1}
=\int_0^{\mu} f(\epsilon) d\epsilon + 2 \sum_{j=1}^{\infty} C_j
\beta^{-2j}
 \ ,
\end{equation} 
where $C_j=\eta(2j) d^{(2j-1)}~f(\epsilon)/d\epsilon^{(2j-1)}|_{\epsilon=\mu}$ and $\eta(x)$ is the Dirichlet eta function.
Using Eqs.(\ref{eq_syst1_N}), (\ref{eq_syst1_E}) and (\ref{eq_dev_sommerfeld}), we define the chemical potential and 
inverse temperature to leading order - $\mu_0$ and $\beta_0$ respectively - as: 
\begin{eqnarray}
N&=&\int_0^{\mu_0} \rho(\epsilon) d\epsilon \ , \label{eq_N_sommer}\\
Q&=&E-E_0= 2\eta(2) \rho(\mu_0) \beta_0^{-2}~,\label{eq_E_sommer}
\end{eqnarray}
where $E_0=\int_0^{\mu_0} \epsilon \rho(\epsilon) d\epsilon$. We have neglected terms in the derivatives of $\rho$ in Eqs.(\ref{eq_N_sommer}) and (\ref{eq_E_sommer}). 
When $\rho$ satisfies Eq.(\ref{eq_rho_power_law}), we can easily invert (\ref{eq_N_sommer}) and (\ref{eq_E_sommer}) 
to obtain:
\begin{equation}\label{eq_mu_approx}
\mu_0=N^{1/\nu}\qquad\text{and}\qquad \beta_0 = \nu \sqrt{2 \eta(2)/Q} N^{(\nu-1)/2\nu}.
\end{equation}
Integrating  the Grand potential Eq.(\ref{def_grd_pot_bos}) by parts, we get:
\begin{equation}\label{eq_o_sommer}
\Omega(\mu_0,\beta_0)=\Omega(0,\mu_0) - 2\eta(2) \rho(\mu_0) \beta_0^{-2}.
\end{equation}
Using the expansions of Eqs.(\ref{eq_N_sommer}),
(\ref{eq_E_sommer}) and (\ref{eq_o_sommer}) in the entropy Eq.(\ref{eq_def_entrop}), the result is
\begin{equation}\label{eq_def_entrop22}
\mathrm{\cal S}(\beta_0,\mu_0) = 
 \beta_0 \bigg[ - \Omega(0,\mu_0) + 2\eta(2) \rho(\mu_0) \beta_0^{-2} + E
-\mu_0 N  \bigg] \ .  
\end{equation}
At zero temperature it leads to:
\begin{equation}
 - \Omega(0,\mu_0) + E_0 -\mu_0 N =0 \ .
\end{equation}
We have used the latter equation in Eq.(\ref{eq_def_entrop22}). Finally we obtain the leading order of the entropy for a Fermi gas at low temperature:
\begin{equation}\label{eq_def_entrop33}
\mathrm{\cal S}(\beta_0,\mu_0) = 2 \sqrt{\frac{\pi^2}{6}\rho(\mu_0)Q} \ ,
\end{equation}
where we have used $\eta(2)=\pi^2/12$. This result holds for any smooth SP level density; in the case Eq.(\ref{eq_rho_power_law}) the entropy is written $\mathrm{\cal S}(Q,N) = 2 \sqrt{\nu \pi^2Q/6}N^{(\nu-1)/2\nu}$. Higher order terms can be easily computed leading to a power series in $Q$ and $N$.
We emphasize that for fermions, the dependence of $\mathrm{\cal S}$ on the excitation energy does not depend on the system. In contrast, its $N$ dependence comes through $\rho(\mu_0)$. This behavior is very different from the bosonic case (except for $\nu=1$, see below) where the energy variation of the entropy strongly depends  on the SP level density, {\it i.e.} $\rhomb \propto \exp[Q^{\nu/(\nu+1)}]$ \cite{tmb,clm}.\\
When the particle number is fixed and when the temperature is sufficiently large, Eq.(\ref{eq_syst1_N}) leads to a large negative chemical potential. $e^{\beta \mu}$ tends to $0$ and the Fermi-Dirac statistic converges to the Maxwell-Boltzmann distribution. In this case Eqs.(\ref{eq_syst2_N}) and (\ref{eq_syst2_E}) become:
\begin{eqnarray}
N&=& \frac{ \Gamma (\nu+1)}{\beta^{\nu}} e^{\beta \mu}
\label{eq_syst6_N} \ , \\ 
E&=& \frac{\nu \Gamma (\nu+1)}{\beta^{\nu+1}} e^{\beta \mu}   \ . \label{eq_syst6_E} 
\end{eqnarray}
We invert Eqs.(\ref{eq_syst6_N}) and (\ref{eq_syst6_E}) to get $\mu$ and $\beta$ as a function of $N$ and $E$. Using Eqs.(\ref{eq_syst1_rho}) and (\ref{eq_def_entrop2}), it yields the following formulas for $E \gg N$: 
\begin{equation}\label{eq_bm}
\rhomb(E,N)= \bigg [ \frac{\Gamma(\nu+1) E^{\nu}}{\nu^{\nu} N^{\nu+1}}
\bigg]^N e^{(\nu+1)N}  \ , 
\end{equation}
and $\mu=\ln[\beta^{\nu} N/\Gamma(\nu+1)]/\beta$.
We also find the usual equation of state for perfect gases: $E=\nu N T$. In contrast to the stretched 
exponential growth at low temperature, we notice that the MB level density has a power-law behavior at high 
temperatures.

%------------------------------------------------------------------------------------------
%------------------------------------------------------------------------------------------
%------------------------------------------------------------------------------------------

\section{The one dimensional harmonic oscillator}\label{1dho}
In this section we compute exactly the MB level density for a system of non-interacting particles confined by
a one-dimensional harmonic potential (1DHO) with unit frequency and make connections with number theory. We choose $\epsilon_0=0$ such that the spectrum is given by integers. We must compute the MB level density for a given integer excitation energy $n$ and for a given particle number $N$, by counting all possible configurations of particles. A schematic way to depict this enumeration consists of using Young diagrams.
They form $n$ boxes with $j$ non-increasing lines such as:
  $c_1 \geq c_2 \geq ...\geq c_j$
, ${\sum_{i=1}^j c_i =n}$ \, where  $c_i$ is the column number for the line $i$ with $j \leq N$. Consequently, for an integer $n$ the number of Young diagrams is the number of ways of counting $n$ as non-zero integers when the line number is less than $N$. This number $p_{_{N}}(n)$ is the so-called {\it partition function of $n$ restricted to at most $N$ terms}.\\
For each Young diagram we associate a unique configuration of particles in such a way that the horizontal axis represents the excitation energy of the particle and the vertical axis represents the index of the occupied level. Hence we have a one-to-one correspondence between each particle configuration and Young diagrams; $p_{_{N}}(n)$ is $\rhomb(n,N)$ \cite{tmb,leboeuf}. In the same way one can easily shows that the result holds for particles with Bose-Einstein statistics as well.\\
The theory of partitions is studied in profusion by mathematicians (see \cite{andrews} for an introduction). One of the most singular results comes from the works of Hardy, Ramanujan and Rademacher who computed an analytical expression when the number of terms into $p_{_{N}}(n)$ is not restricted anymore \cite{hr,r}. In this case $p_{_{N \geq n}}(n)=p(n)$ is called the {\it partition number of an integer $n$}.
The leading asymptotic expression of $p(n)$ for large $n$ is given by
\begin{equation}\label{eq_form_asymp}
p(n) \approx \frac{1}{\sqrt{48 } n} e^{ \sqrt{2 \pi^2 n/ 3}} \ .
\end{equation}
To leading order the exponential term in Eq.(\ref{eq_form_asymp}) is equivalent to Eq.(\ref{eq_syst1_rho}) with Eq.(\ref{eq_def_entrop33}) and $\nu=1$ as expected.
The correction term for a finite number of terms to $p_{_{N}}(n)$ was computed by Erd\"os and Lehner \cite{el}:
\begin{equation}\label{gumbel}
\frac{p_{_{N}}(n)}{p(n)} \approx e^{- e^{- g}} ,
%p_{_{N}}(n)/p(n) \approx e^{- e^{- g}} ,
\end{equation}
with $g=\sqrt{\pi^2/(6n)}N + 1/2 \ln[\pi^2/(6n)]$ and $N=o(n^{1/3})$. This distribution is the so-called Gumbel-like, and is one of the universal distributions in the domain of extreme values statistics \cite{coles}.
It shows how the finite particle number contributes to the decrease of the number of configurations when heating the system (increases $n$) and determines the transition from low- to high-temperature regime. The main target of the article is to evaluate this correction when the spectrum is no longer equidistant but fulfills the more general
law Eq.(\ref{eq_rho_power_law}). This will be computed in the following section.

Note that in the case of fermions, the Fermi energy depends on the particle number, while this is not the case for bosons. This fundamental difference implies that the equivalence between the two statistics is only valid in the case of the 1DHO. In the context of the partition theory the generalization to a SP level density of the form Eq.(\ref{eq_rho_power_law}) leads to map $\rhomb(E,N)$ to the {\it partition number of an integer $n=E$ into a sum of $N$} distinct (due to Pauli principle) {\it $1/\nu$-th powers of integers}. For bosons, a similar relation exits but without any distinction between summands \cite{tmb,clm}.

%------------------------------------------------------------------------------------------
%------------------------------------------------------------------------------------------
%------------------------------------------------------------------------------------------
\section{General case}\label{gc}
In the degenerate gas limit the chemical potential is large with respect to the excitation energy. At low temperature, the fugacity $z=e^{\beta \mu}$ tends to infinity. To solve the equations Eqs.(\ref{eq_syst2_N}) and (\ref{eq_syst2_E}), we must find the asymptotic expansion of the polylogarithm function for large negative argument.
Garoni {\it et al.} give the inversion expression of $\Li_{\nu}(-z)$ \cite{gfg}: 
\begin{equation}\label{eq_Li_inv}
\Li_{\nu}(-z)  =  - 2 \sum_{n=0}^{\infty}
\frac{\eta(2 n) \ln^{\nu- 2 n} (z^{-1})}{\Gamma(\nu+1-2 n)}
- \cos(\pi \nu) \Li_{\nu}(-z^{-1})  \ .
\end{equation}
Note that for $\nu$ half-integer, the term in $\Li_{\nu}(-z^{-1})$ vanishes. Furthermore when $\nu$ is an integer, the sum in Eq.(\ref{eq_Li_inv}) tends to the integer part of $\nu/2$.
At low temperature $z^{-1}$ tends to zero and $\displaystyle \lim_{x \to 0} \Li_{\nu}(x)=x$. Eqs.(\ref{eq_syst2_N}) and (\ref{eq_syst2_E}) with Eq.(\ref{eq_Li_inv}) become:
\begin{eqnarray}
N= 2 \ \Gamma(\nu+1) \sum_{n=0}^{\infty} \frac{\eta(2
n) \beta^{-2 n} \mu^{\nu - 2 n}}{\Gamma(\nu+1-2 n)} \nonumber\\
 \hspace{1.cm}-\cos(\pi \nu) \frac{\Gamma(\nu+1) \ e^{-\beta \mu}}{\beta^{\nu}} \label{eq_syst3_N}
\end{eqnarray}
and
\begin{eqnarray}
E = 2 \ \nu  \Gamma(\nu+1) \sum_{n=0}^{\infty} \frac{\eta(2
n)\beta^{-2 n} \mu^{\nu +1 - 2 n}}{\Gamma(\nu+2-2 n)} \nonumber\\
\hspace{1.cm}+ \cos(\pi \nu) \frac{\nu \Gamma(\nu+1) \ e^{-\beta
\mu}}{\beta^{\nu+1}} \ .
\label{eq_syst3_E}
\end{eqnarray}
%------------------------------------------------------------------------------------------
%------------------------------------------------------------------------------------------
%------------------------------------------------------------------------------------------
\begin{figure}[ht]
\begin{center}
\includegraphics[width=1.\columnwidth,clip=true]{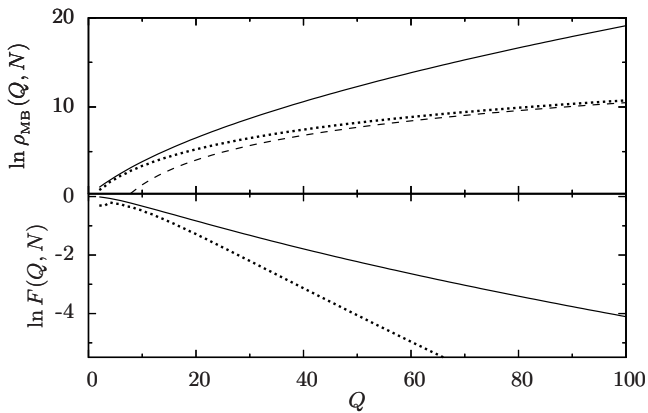} 
\caption{\label{fig} 
Upper panel: The MB level density for $\nu=1$ and for $5$ fermions as a function of the excitation energy $Q$. The dotted line is the exact computation. The full line corresponds to the degenerate or low temperature regime, Eq.~(\ref{eq_form_asymp}) (with $n=Q$). The dashed line corresponds to the high temperature regime, Eq.~(\ref{eq_bm}) (with $E\approx Q$; we have included the expression of the determinant ${\cal D}=E^2$ [see Eq.(\ref{eq_syst1_rho}) and below for discussion]). Lower panel: The function $F(Q,N)$ for the same system. The dotted line corresponds to the exact computation and the full line to the analytical expression Eq.(\ref{lamb_ferm}).}
\end{center}
\end{figure}
%------------------------------------------------------------------------------------------
%------------------------------------------------------------------------------------------
%------------------------------------------------------------------------------------------
Terms in the sum on the r.h.s. of Eqs.(\ref{eq_syst3_N}) and (\ref{eq_syst3_E}) correspond to the Sommerfeld expansion terms. The meaning of the exponential term in Eqs.(\ref{eq_syst3_N}) and (\ref{eq_syst3_E}) will emerge from the following example. With $\nu=1$, Eqs.(\ref{eq_syst3_N}) and (\ref{eq_syst3_E}) yield:
\begin{eqnarray}
N&=& \mu + \frac{e^{- \beta \mu}}{\beta} \ ,
\label{eq_syst7_N} \\
E &=& \frac{\mu^2}{2} + \frac{2 \eta(2)}{\beta^2} - \frac{e^{- \beta \mu}}{\beta^2}  \ . 
\label{eq_syst7_E}
\end{eqnarray}
The computation of $\cal S$ to the second order is done by expanding $\mu$ and $\beta$ in Eqs.(\ref{eq_syst7_N}) and (\ref{eq_syst7_E}). We have:
\begin{eqnarray}
\mu&=&\mu_0- \sqrt{\frac{Q}{2 \eta(2)}} e^{ -\sqrt{2 \eta(2) /Q}
N} \label{eq_mu_nu=1} \ ,\\
\beta &=& \beta_0- \frac{\bigg [N+\sqrt{\frac{Q}{2 \eta(2)}}\bigg]}{2 Q} e^{ -\sqrt{2 \eta(2) /Q}N}  \ . \label{eq_beta_nu=1}
\end{eqnarray}
Using Eqs.(\ref{eq_syst7_N}), (\ref{eq_syst7_E}),  (\ref{eq_mu_nu=1}) and (\ref{eq_beta_nu=1}) into Eq.(\ref{eq_def_entrop2}), we get 
\begin{equation} \label{entrop_nu=1}
\mathrm{\cal S} = 2 \sqrt{2 \eta(2) Q} -e^{-t} \ ,
\end{equation}
where $t= \sqrt{2 \eta(2) /Q}N + \ln [\sqrt{2 \eta(2) /Q}]$. Subtracting the leading order entropy Eq.(\ref{eq_def_entrop33}) to Eq.(\ref{entrop_nu=1}) and using Eqs.(\ref{eq_syst1_rho}) and (\ref{eq_rho_power_law}), Eq.(\ref{gumbel}) is found with $n=Q$. The Gumbel law is thus recovered, as already stated in the previous section. The main contributions to $\rhomb$ are given by the Sommerfeld terms. These terms depend only on the local properties of the SP level density close to the Fermi energy, and ignore the fact that the single particle spectrum is bounded
from below. The transition to the Maxwell-Boltzmann regime, where all particles contribute to a single configuration, cannot be described by this expansion.
The exponentially small term $e^{-\beta \mu}$ in Eqs.(\ref{eq_syst3_N}) and (\ref{eq_syst3_E}) takes into account this effect, and this is the reason why we call it a {\it non-local} correction to the MB level density.
For a given $\nu$, we compute the non-local contribution to $\rhomb$ using the previous method, to obtain 
%\vspace{-1.cm}
\begin{eqnarray}
F(Q,N)&= &\frac{\rhomb(Q,N)}{\overline{\rhomb}(Q,N)} \nonumber \\
&=&\exp \bigg[\frac{\cos(\pi
\nu) \Gamma(\nu+1) }{\beta_0^{\nu}} \exp(-\beta_0 \mu_0)\bigg] \ .~~~~
\label{lamb_ferm}
\end{eqnarray}
$\overline{\rhomb}(Q,N)$ represents the smooth MB level density computed by means of the (local) Sommerfeld expansion Eq.(\ref{eq_dev_sommerfeld}). Equation (\ref{lamb_ferm}) is the main result of the paper.
The final correction takes the form of a modified Gumbel law. This is different with respect to the case of bosons,
where, depending on the value of $\nu$, three different functional dependencies of the correction were found \cite{clm}.
They correspond to the three different distributions that appear in the theory of extreme value statistics. The difference is related, in particular, to the fact that there is no Bose-Einstein condensation for non-interacting fermions. 

The qualitative behavior of $F(Q,N)$ has a strong dependence on the value of $\nu$. For instance, 
$F(Q,N)=1$ for $\nu$ half-integers. It can be larger than one for some specific values of $\nu$. 
This effect is not surprising, since for $\nu\neq1$ the Sommerfeld expansion gives the leading variation in $N$. In that case $F(Q,N)$ is not constraint to be less than one. 
The behavior of the MB level density as a function of the excitation energy is illustrated in Fig.~\ref{fig} for $\nu=1$ and for $N=5$ fermions. This value of $\nu$ describes a one-dimensional harmonic oscillator, but also for instance a two-dimensional billiard of arbitrary shape. A low number of particles has been chosen in order to enhance the finite $N$ cotrrections. The upper panel illustrates the transition from the degenerate gas regime to the Maxwell-Boltzmann one. The lower panel shows the correction factor $F(Q,N)$. We note that the correction term is significant only in a small region around the degenerate regime, {\it i.e.} for $Q\lesssim 15$. At higher excitation energies, the Maxwell-Boltzmann regime is reached.

In conclusion we have demonstrated that in the transition from the low to the high temperature regimes the MB level density of a Fermi gas is described by a modified law of extreme value statistics. This result is valid for a large set of systems according to Eq.(\ref{eq_rho_power_law}). In the context of number theory, our results apply in particular to the partition of integers as a sum of powers of distinct integers. Our results are of interest in the computation of the level density of small nuclei for which the finite size correction term (\ref{lamb_ferm}) is the most important.\\
J. R. acknowledges M. V. N. Murthy for fruitful discussions and C. Torrero for his careful reading of the manuscript. 
J. R. thanks also financial support from the French National Research Agency ANR (project ANR-06-BLAN-0059).
%------------------------------------------------------------------------------------------
%------------------------------------------------------------------------------------------
%-----------------------------------------------------------------------------------------


\begin{thebibliography}{10}

\bibitem{nuc1} N. Rosenzweig. Phys. Rev. {\bf 108}, 817 (1957).

\bibitem{nuc2} A. Bohr and B. R. Mottelson, {\it Nuclear Structure I.} (W.A. Benjamin, New York 1969).

\bibitem{bh1} A. Ashtekar, J. Baez, A. Corchini and K. Krasnov, Phys. Rev. Lett. {\bf 80}, 904 (1998).

\bibitem{bh2} S. Das, P. Majumdar and R. K. Bhaduri, Classical Quantum Grav. {\bf 19}, 2355 (2002).

\bibitem{bh3} A. Alsleev, A. P. Polychronakos, M. Smedib\"ack, Phys. Lett. B {\bf 574}, 296 (2003). 
 
\bibitem{bethe} H. A. Bethe, Phys. Rev. {\bf 50}, 332 (1936).

\bibitem{dsvu} W. Dilg, W. Schantl, H. Vonach and M. Uhl, Nucl.Phys. A {\bf 217} 269 (1973).

\bibitem{gc} A. Gilbert, A. G. W. Cameron, Can. J. Phys. {\bf 43}, 1446 (1965).

\bibitem{clm} A. Comtet, P. Leboeuf, S. N. Majumdar, Phys. Rev. Lett. {\bf 98}, 070404 (2007).

\bibitem{tmb} M. N. Tran, M. V. N. Murthy, and R. K. Bhaduri, Ann. Phys. (NY) {\bf 311}, 204 (2004).

\bibitem{leboeuf} P. Leboeuf, AIP Conference Proceedings Vol. {\bf 777} 180 (2005)  V. Zelevinsky (ed.). 

\bibitem{gfg} T. M. Garoni, N. E. Frankel, M. L. Glasser, J. Math. Phys. {\bf 42}, 1860 (2001). 

\bibitem{s} A. Sommerfeld, Zeit. f. Phys. A {\bf 47}, 1 (1928).

\bibitem{am} N. W. Ashcroft, N. D. Mermin, {\it Solid State Physics} (Sauders College Publishing, USA, 1976).

\bibitem{andrews} G. E. Andrews, {\sl The Theory of Partitions} (Cambridge University Press, Cambridge, 1998).

\bibitem{hr} G. H. Hardy and S. Ramanujan, Proc. London Math. Soc. {\bf 17}, 75 (1918).

\bibitem{r} H. Rademacher, Proc. London Math. Soc. {\bf 43}, 241 (1937).

\bibitem{el} P. Erd\"os and J. Lehner, Duke Math J. {\bf 8}, 335 (1941).

\bibitem{coles} S. Coles, {\it An introduction to statistical modeling of extreme values} (Springer, 2001).


\end{thebibliography}
\end{document}